\documentclass[]{spie}  

\usepackage{xcolor}
\usepackage{amsmath,amsfonts,amssymb}
\usepackage{graphicx}
\usepackage[colorlinks=true, allcolors=blue]{hyperref}
\usepackage{siunitx}
\usepackage{makecell}
\usepackage[acronym]{glossaries}
\usepackage[nolist]{acronym}
\newacro{SAF}[SAF]{Science Archive Facility}
\newacro{SOXS}[SOXS]{Son-Of-X-shooter}
\newacro{NTT}[NTT]{New Technology Telescope}
\newacro{ESO}[ESO]{European Southern Observatory}
\newacro{DRAWER}[DRAWER]{Data Reduction And WEllness Reporter}
\newacro{AC}[AC]{Acquisition Camera}
\newacro{CPL}[CPL]{Common Pipeline Library}
\newacro{CLI}[CLI]{Command-Line Interface}
\newacro{QC}[QC]{Quality Control}

\usepackage{color}
\usepackage{colortbl}
\usepackage{multirow}
\usepackage{listings}
\usepackage{float}
\usepackage{color}
\usepackage{xcolor,colortbl}
\definecolor{red}{HTML}{dc322f}
\definecolor{green}{HTML}{859900}
\definecolor{blue}{HTML}{268bd2}
\definecolor{orange}{HTML}{cb4b16}
\definecolor{cyan}{HTML}{2aa198}
\definecolor{magneta}{HTML}{d33682}
\definecolor{violet}{HTML}{6c71c4}

\definecolor{dkgreen}{rgb}{0,0.6,0}
\definecolor{gray}{rgb}{0.5,0.5,0.5}
\definecolor{mauve}{rgb}{0.58,0,0.82}

\title{The SOXS Data-Reduction Pipeline}

\author[a]{David R. Young}
\author[b]{Marco Landoni}
\author[a]{Stephen J. Smartt}

\author[b]{Sergio~Campana}
\author[s]{Riccardo~Claudi}
\author[d]{Pietro~Schipani}
\author[b]{Matteo~Aliverti}
\author[s]{Andrea~Baruffolo}
\author[e]{Sagi~Ben-Ami}
\author[a,i]{Federico~Biondi}
\author[d]{Giulio~Capasso}
\author[f]{Rosario~Cosentino}
\author[g]{Francesco~D'Alessio}
\author[c]{Paolo~D'Avanzo}
\author[h]{Ofir	Hershko}
\author[j,q]{Hanindyo~Kuncarayakti}
\author[k]{Matteo~Munari}
\author[m,t]{Giuliano~Pignata}
\author[h]{Adam~Rubin}
\author[k]{Salvatore~Scuderi}
\author[g]{Fabrizio~Vitali}
\author[l]{Jani~Achrén}
\author[m,t]{José~Antonio~Araiza-Duran}
\author[n]{Iair~Arcavi}
\author[c]{Anna~Brucalassi}
\author[h]{Rachel~Bruch}
\author[a]{Enrico~Cappellaro}
\author[d]{Mirko~Colapietro}
\author[d]{Massimo~Della~Valle}
\author[s]{Marco~De~Pascale}
\author[k]{Rosario~Di~Benedetto}
\author[d]{Sergio~D'Orsi}
\author[h]{Avishay~Gal-Yam}
\author[b]{Matteo~Genoni}
\author[f]{Marcos~Hernandez}
\author[j,q]{Jari~Kotilainen}
\author[r]{Gianluca~Li~Causi}
\author[q]{Seppo~Mattila}
\author[h]{Michael~Rappaport}
\author[s]{Kalyan~Radhakrishnan}
\author[s]{Davide~Ricci}
\author[b]{Marco~Riva}
\author[s]{Bernardo~Salasnich}
\author[k]{Ricardo~Zanmar~Sanchez}
\author[u]{Maximilian~Stritzinger}
\author[f]{Hector~Ventura}

\affil[a]{Astrophysics Research Centre, School of Mathematics and Physics, Queen's University Belfast, Belfast BT7 1NN, UK}
\affil[b]{INAF -- Osservatorio Astronomico di Brera, Via Bianchi 46, I-23807, Merate, Italy }
\affil[c]{ESO, Karl Schwarzschild Strasse 2, D-85748, Garching bei München, Germany }
\affil[d]{INAF -- Osservatorio Astronomico di Capodimonte, Sal. Moiariello 16, I-80131, Naples, Italy }
\affil[e]{Harvard-Smithsonian Center for Astrophysics, Cambridge, USA }
\affil[f]{FGG-INAF, TNG, Rambla J.A. Fernández Pérez 7, E-38712 Breña Baja (TF), Spain }
\affil[g]{INAF -- Osservatorio Astronomico di Roma, Via Frascati 33, I-00078 M. Porzio Catone, Italy }
\affil[h]{Weizmann Institute of Science, Herzl St 234, Rehovot, 7610001, Israel }
\affil[i]{Max-Planck-Institut für Extraterrestrische Physik, Giessenbachstr. 1, D-85748 Garching, Germany }
\affil[j]{Finnish Centre for Astronomy with ESO (FINCA), FI-20014 University of Turku, Finland}
\affil[k]{INAF -- Osservatorio Astrofisico di Catania, Via S. Sofia 78 30, I-95123 Catania, Italy }
\affil[l]{Incident Angle Oy, Capsiankatu 4 A 29, FI-20320 Turku, Finland }
\affil[m]{Universidad Andres Bello, Avda. Republica 252, Santiago, Chile }
\affil[n]{Tel Aviv University, Department of Astrophysics, 69978 Tel Aviv, Israel }
\affil[o]{Dark Cosmology Centre, Juliane Maries Vej 30, DK-2100 Copenhagen, Denmark }
\affil[p]{Aboa Space Research Oy, Tierankatu 4B, FI-20520 Turku, Finland}
\affil[q]{Tuorla Observatory, Dept. of Physics and Astronomy, FI-20014 University of Turku, Finland }
\affil[r]{INAF - Istituto di Astrofisica e Planetologia Spaziali, Rome, Italy}
\affil[s]{INAF -- Osservatorio Astronomico di Padova, Vicolo dell’Osservatorio 5, I-35122, Padua, Italy }
\affil[t]{Millennium Institute of Astrophysics (MAS)}
\affil[u]{Aarhus University, Ny Munkegade 120, D-8000 Aarhus, Denmark }

\authorinfo{Further author information: (Send correspondence to David Young)\\David Young: E-mail: d.r.young@qub.ac.uk\\  Marco Landoni: E-mail: marco.landoni@inaf.it\\Stephen Smartt: E-mail:  s.smartt@qub.ac.uk}

\begin{document} 
\maketitle

\begin{abstract}
The \ac{SOXS} is a dual arm spectrograph (UV-VIS \& NIR) and \ac{AC} due to mounted on the \ac{ESO} 3.6m \ac{NTT} in La Silla. Designed to simultaneously cover the optical and NIR wavelength range from 350-2050 nm, the instrument will be dedicated to the study of transient and variable events with many Target of Opportunity requests expected.

The goal of the \ac{SOXS} Data Reduction pipeline is to use calibration data to remove all instrument signatures from the \ac{SOXS} scientific data frames for each of the supported instrument modes, convert this data into physical units and deliver them with their associated error bars to the \ac{ESO} \ac{SAF} as Phase 3 compliant science data products, all within 30 minutes. The primary reduced product will be a detrended, wavelength and flux calibrated, telluric corrected 1D spectrum with UV-VIS + NIR arms stitched together. The pipeline will also generate \ac{QC} metrics to monitor telescope, instrument and detector health.

The pipeline is written in Python 3 and has been built with an agile development philosophy that includes adaptive planning and evolutionary development. The pipeline is to be used by the \ac{SOXS} consortium and the general user community that may want to perform tailored processing of \ac{SOXS} data. Test driven development has been used throughout the build using `extreme' mock data. We aim for the pipeline to be easy to install and extensively and clearly documented.
\end{abstract}


\section{INTRODUCTION}
\label{sec:intro}  

The SOXS (Son Of X-Shooter) instrument is a new medium resolution spectrograph ($R\simeq4500$) capable of simultaneously observing 350-2000nm (U- to H-band) to a limiting magnitude of R $\sim 20$ (3600sec, S/N $\sim$ 10). It shall be hosted at the Nasmyth focus of the \ac{NTT} at La Silla Observatory, Chile (see \cite{soxsschipani} for an overview). This paper describes the design of the \ac{SOXS} data-reduction pipeline and data-products it shall generate and is part of a series of contributions \cite{soxsschipani,
  soxslandoni, soxsrubin, soxsbiondi, soxsgenoni,
  soxskuncarayakti, soxsyoung, soxsbrucalassi, soxscolapietro,
  soxscosentino, soxsclaudi, soxsaliverti, soxssanchez, soxsvitali}
describing the current development status of the SOXS subsystems.

Section \ref{sec:detectors} gives details about the individual instruments that the \ac{SOXS} pipeline shall receive data from. Section \ref{sec:obsmodes} details the various observation modes \ac{SOXS} will be able to operate in and the products the pipeline is expected to generate for each of those modes. In Section \ref{sec:automated_data_reduction} we propose how we will automate much of the data-reductions for \ac{SOXS}, followed by how users in will be able to interact with the pipeline on their own machines in Section \ref{sec:usage}, the development environment we are using in Section \ref{sec:development_environment} and finally giving an overview of the architecture of the pipeline code-base in Section \ref{sec:architecture}.

\section{Three SOXS Detectors}
\label{sec:detectors} 

\begin{figure}[h]
  \begin{center}
  \begin{tabular}{c} 
  \includegraphics[width=10cm]{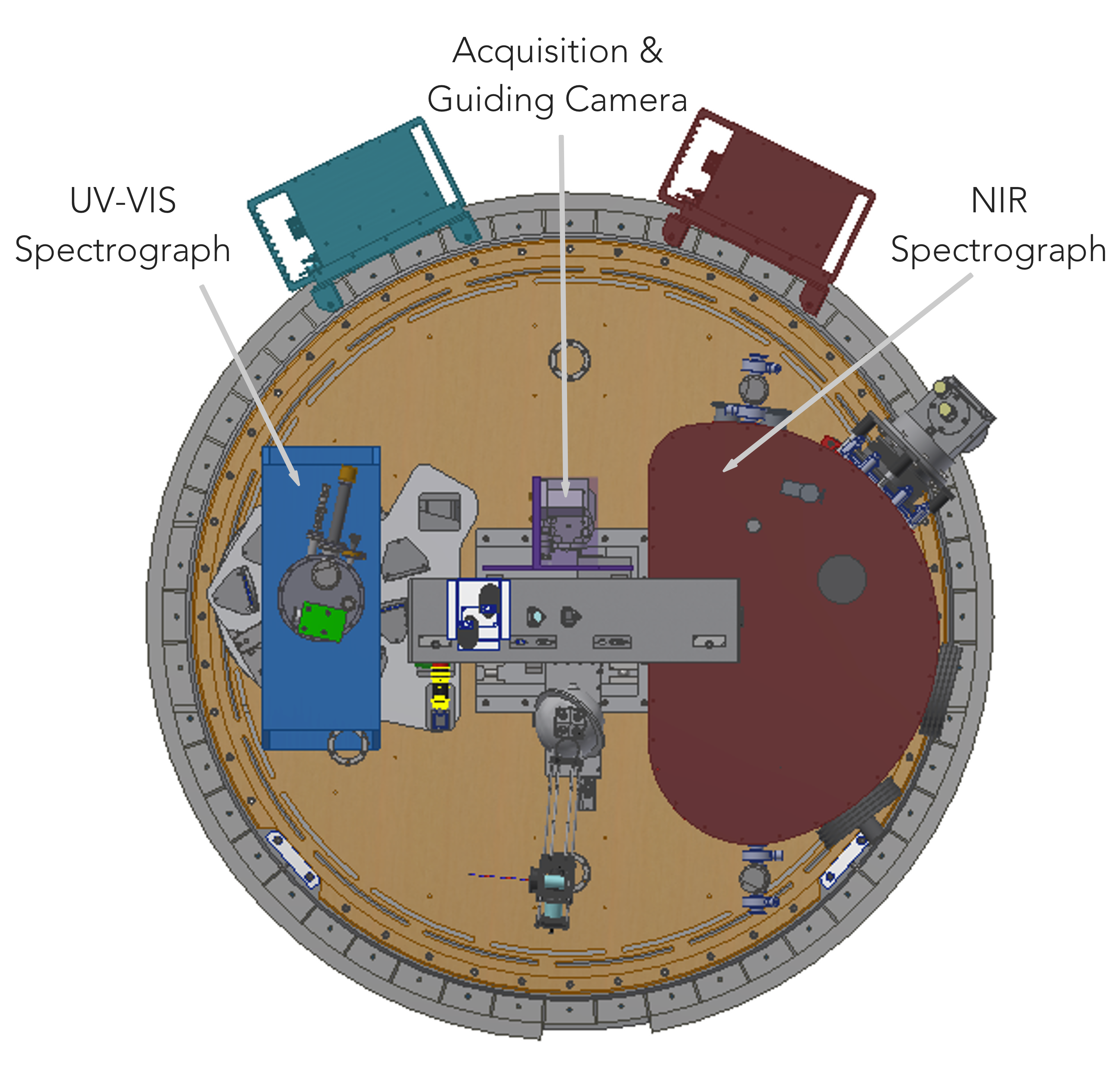}
  \end{tabular}
  \end{center}
  \caption[soxs-on-flange] 
  { \label{fig:soxs-on-flange} 
A face-on view of \ac{SOXS} on the \ac{NTT} nasmyth-focus rotator flange.}
  \end{figure}

SOXS will host three instruments; the UV-VIS and NIR spectrographs and the \ac{AC}. Figure \ref{fig:soxs-on-flange} shows how these instruments are to be mounted on the NTT's Nasmyth focus rotator-flange. The \ac{SOXS} pipeline will be capable of reducing the pixel-data collected by each of these instruments as well as outputting Quality Control metrics used to monitor the health of each unit. 

\subsection{UV-VIS Spectrograph}

  \begin{figure}[h]
  \begin{center}
  \begin{tabular}{c} 
  \includegraphics[width=10cm]{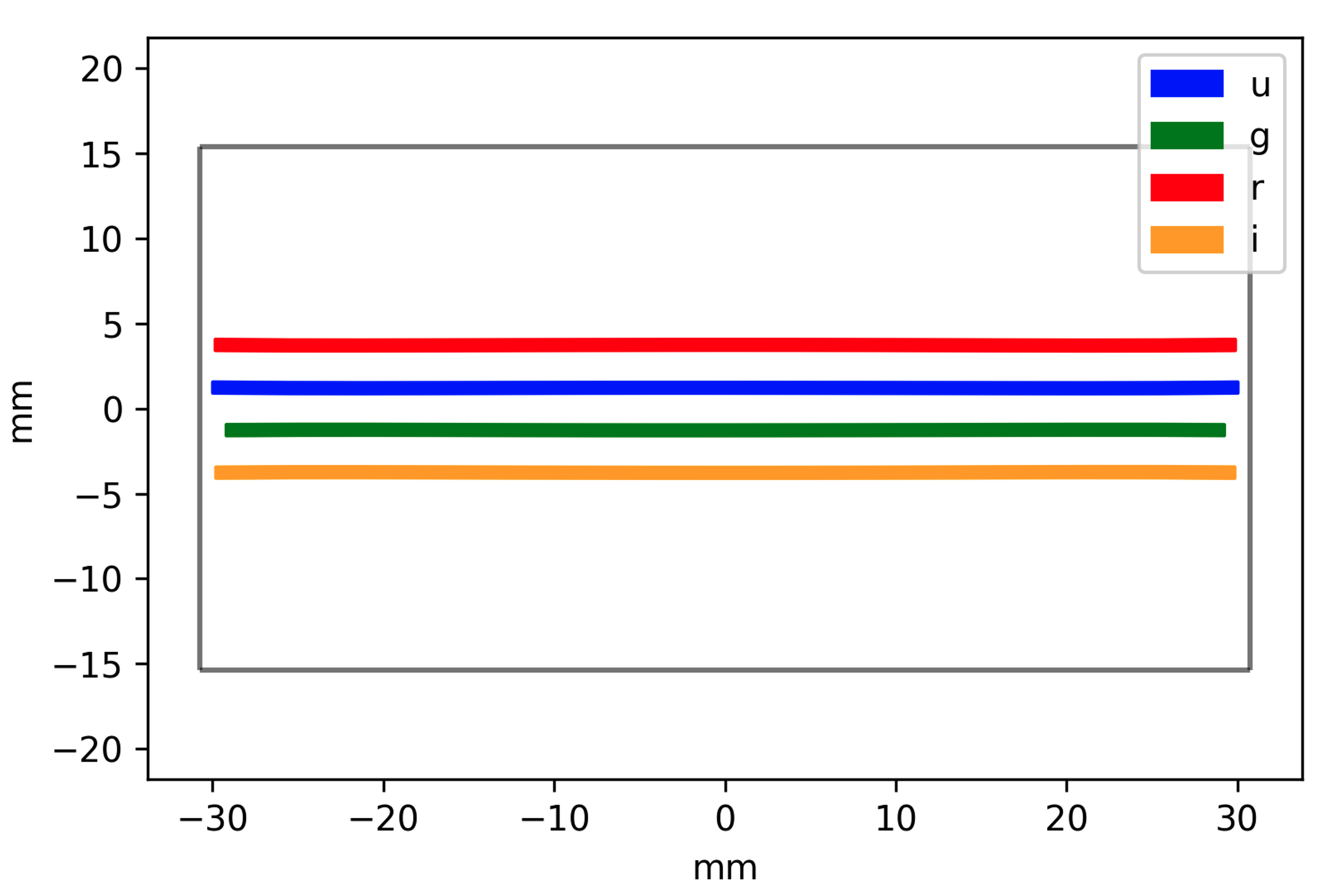}
  \end{tabular}
  \end{center}
  \caption[uvvis-spectral-format] 
  { \label{fig:uvvis-spectral-format} 
The four quasi-orders of the \ac{SOXS} UV-VIS spectral format realised on the UV-VIS detector plain.}
  \end{figure} 

The UV-VIS spectrograph \cite{soxsrubin} employs 4 ion-etched transmission gratings in the first order ($m=1$) to record spectra in the 350-850nm wavelength range (providing a 50nm overlap with the NIR arm for cross-calibration). The spectral band is split into four poly-chromatic channels and sent to their specific grating. The spectral format of these four quasi-orders will be linear as shown in  Figure \ref{fig:uvvis-spectral-format}. The main characteristics of the UV-VIS spectrograph and the e2V CCD44-82 detector \cite{soxscosentino} can be found in Table \ref{tab:uvvis-spectrograph-ccd-characteristics}.

\begin{table}[ht]
\caption{Characteristics of the \ac{SOXS} UV-VIS Spectrograph \& CCD.} 
\label{tab:uvvis-spectrograph-ccd-characteristics}
\begin{center}       
\begin{tabular}{|l|l|}
\hline
\rule[-1ex]{0pt}{3.5ex} Detector            & e2V CCD44-82                                       \\ \hline
\rule[-1ex]{0pt}{3.5ex} Pixel-Size          & \SI{15}{\micro\metre}                                          \\ \hline
\rule[-1ex]{0pt}{3.5ex} Array-Size          & 2048$\times$4096px; 30.7$\times$61.4mm              \\ \hline
\rule[-1ex]{0pt}{3.5ex} Array-Scale         & 0.28 arcsec/px                                     \\ \hline
\rule[-1ex]{0pt}{3.5ex} Peak Signal         & 200,000 e$^-$/px                                    \\ \hline
\rule[-1ex]{0pt}{3.5ex} Gain                & \makecell[l]{Slow: 0.6 ± 0.1 e$^-$/ADU  \\ Fast: 2 ± 0.2 e$^-$/ADU } \\ \hline
\rule[-1ex]{0pt}{3.5ex} Read noise (rms)    & \makecell[l]{Slow: $<$ 3e$^-$ \\ Fast: $<$8 e$^-$ }  \\ \hline
\rule[-1ex]{0pt}{3.5ex} Dark current @ 153K &  $<$0.00001 e$^-$/s/px                                \\ \hline
\rule[-1ex]{0pt}{3.5ex} Resolution (R)      & 3500-7000 ($\simeq$ 4500 mean)                     \\ \hline
\rule[-1ex]{0pt}{3.5ex} Wavelength Range    & 350-850nm                                          \\ \hline
\rule[-1ex]{0pt}{3.5ex} Slit Widths         & 0.5, 1.0, 1.5, 5.0 arcsec                          \\ \hline
\rule[-1ex]{0pt}{3.5ex} Slit Length         & 12 arcsec                                          \\ \hline
\rule[-1ex]{0pt}{3.5ex} Grating Blaze Angle & 41°                                                \\ \hline
\rule[-1ex]{0pt}{3.5ex} Orders (quasi)      & 4                                                  \\ \hline
\end{tabular}
\end{center}
\end{table}
%

\subsection{NIR Spectrograph}

\begin{figure} [ht]
  \begin{center}
  \begin{tabular}{c} 
  \includegraphics[width=10cm]{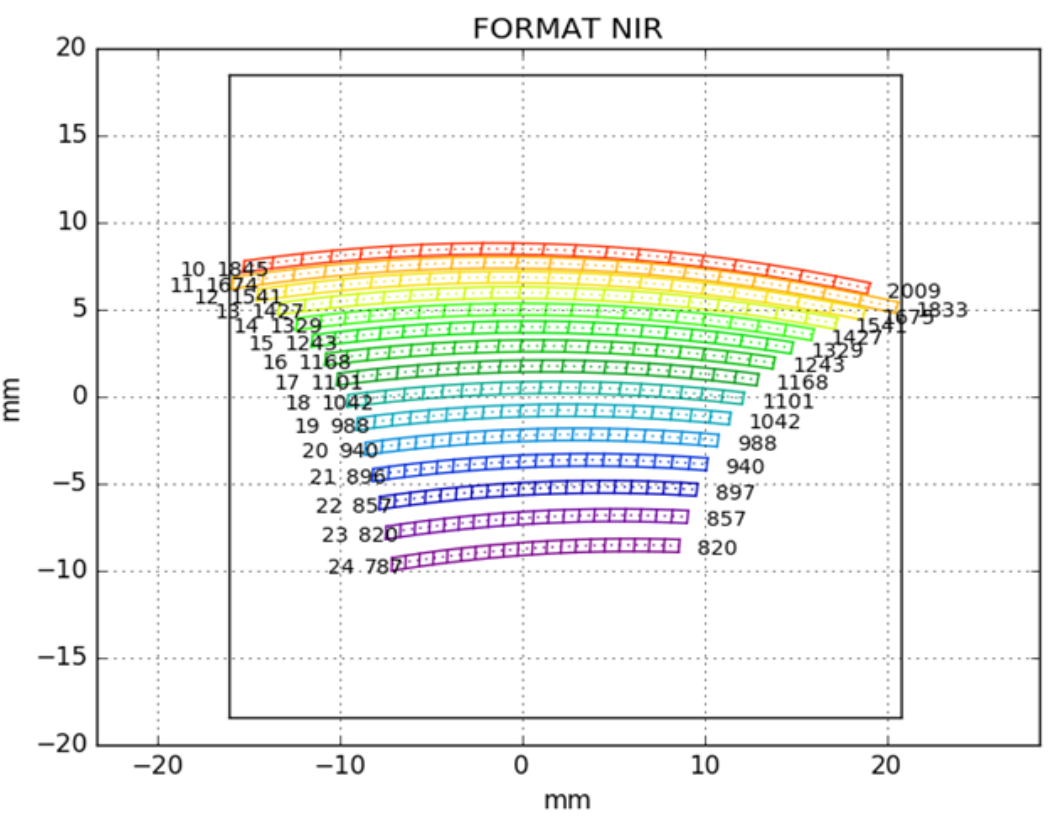}
  \end{tabular}
  \end{center}
  \caption[nir-spectral-format] 
  { \label{fig:nir-spectral-format} 
The 15 orders of the \ac{SOXS} NIR spectral format realised on the NIR array plain}
  \end{figure}

The \ac{SOXS} NIR spectrograph \cite{soxsvitali} is a cross-dispersed echelle that employs the `4C' (Collimator Correction of Camera Chromatism) to obtain spectra in 800-2000nm wavelength range in 15 orders (providing a 50nm overlap with the UV-VIS arm for cross-calibration). Unlike the UV-VIS spectrograph the NIR orders display a curvature as evidenced by the spectral format of the NIR orders shown in Figure \ref{fig:nir-spectral-format}.

\begin{table}[ht]
\caption{Characteristics of the \ac{SOXS} NIR Spectrograph \& Detector Array} 
\label{tab:nir-spectrograph-array-characteristics}
\begin{center} 
\begin{tabular}{|l|l|}
\hline
\rule[-1ex]{0pt}{3.5ex} Detector                    & Teledyne H2RG                                            \\ \hline
\rule[-1ex]{0pt}{3.5ex} Pixel-Size                  & \SI{18}{\micro\metre}                                                    \\ \hline
\rule[-1ex]{0pt}{3.5ex} Array-Size                  & 2048$\times$2048px                                    \\ \hline
\rule[-1ex]{0pt}{3.5ex} Array-Scale                 & 0.25 arcsec/px                                           \\ \hline
\rule[-1ex]{0pt}{3.5ex} Read noise (rms)            & \makecell[l]{Double correlated: $<$20 e$^-$ \\ 16 Fowler pairs $<$7 e$^-$} \\ \hline
\rule[-1ex]{0pt}{3.5ex} Dark current @ 40K          & $<$0.005 e$^-$/s/px                                        \\ \hline
\rule[-1ex]{0pt}{3.5ex} Resolution (R)              & $\simeq$ 5000 (1 arcsec slit)                            \\ \hline
\rule[-1ex]{0pt}{3.5ex} Wavelength Range            & 800-2000 nm                                              \\ \hline
\rule[-1ex]{0pt}{3.5ex} Slit Widths                 & 0.5, 1.0, 1.5, 5.0 arcsec                                \\ \hline
\rule[-1ex]{0pt}{3.5ex} Slit Length                 & 12 arcsec                                                \\ \hline
\rule[-1ex]{0pt}{3.5ex} Grating Blaze Angle         & 44°                                                      \\ \hline
\rule[-1ex]{0pt}{3.5ex} Detector Operating Temp     & 40K                                                      \\ \hline
\rule[-1ex]{0pt}{3.5ex} Spectrograph Operating Temp & 150K                                                     \\ \hline
\rule[-1ex]{0pt}{3.5ex} Orders                      & 15                                                       \\ \hline
\end{tabular}
\end{center}
\end{table}

\subsection{Acquisition Camera}

The Acquisition Camera \cite{soxsbrucalassi} with a $3.5' \times 3.5'$ FOV will also allow for science-grade multi-band photometry.

The primary use of the acquisition camera is to acquire spectral targets and centre them on the slits. In addition, the camera's $3.5' \times 3.5'$ FOV and 0.205 arcsec/px scale will also allow for science-grade multi-band photometry. Observers will be able to select from 7 filters; the LSST $u, g, r, i, z, y$ set and Johnson $V$.

\begin{table}[ht]
\caption{Characteristics of the \ac{SOXS} \ac{AC}.} 
\label{tab:agcam-specs}
\begin{center}       
\begin{tabular}{|l|l|l|}
\hline
\rule[-1ex]{0pt}{3.5ex} Camera & Andor iKon M934  \\
\hline
\rule[-1ex]{0pt}{3.5ex}  Detector & BEX2-DD   \\
\hline
\rule[-1ex]{0pt}{3.5ex}  Pixel-Size & \SI{13}{\micro\metre}  \\
\hline
\rule[-1ex]{0pt}{3.5ex}  Array-Size &  1024$\times$1024;  13.3$\times13.3$mm  \\
\hline
\rule[-1ex]{0pt}{3.5ex}  Peak Signal & 130000 e$^-$/px  \\
\hline 
\rule[-1ex]{0pt}{3.5ex}  Dark Current @ 173 K & 0.00012 e$^-$/s/px  \\
\hline
\rule[-1ex]{0pt}{3.5ex}  Read noise (rms) & 2.9 e$^-$  \\
\hline
\rule[-1ex]{0pt}{3.5ex}  Filters & \emph{u, g, r, i, z, y, V}  \\
\hline
\end{tabular}
\end{center}
\end{table}

\section{Observation Modes and Pipeline Data-Products}
\label{sec:obsmodes} 

The challenge faced by any pixel-based data reduction pipeline is to
identify and remove (or minimise) all sources of noise from the raw data
so as to be able to best extract the scientific information contained on
the frames. Alongside the standard detrending stages of calibration
needed to remove instrument signatures (bias, dark-removal and
flat-field correcting), the pipeline will calculate and apply
accurate wavelength- and flux-calibration solutions to the spectra. For
observations taken in the standard stare-mode it must also solve the
notoriously difficult problem of removing the unwanted signal of diffuse
light scattered in the earth's atmosphere (the \emph{sky-background}).
Finally, the pipeline must provided the means to extract the object
spectra from the 2D image frames in a manner that maximises the
signal-to-noise of the data; otherwise known as
\emph{optimal-extraction}.

SOXS will be able to operate in 6 different spectroscopic observation modes:

\begin{enumerate}
\item
  \textbf{Stare-mode}. Standard `point-and-shoot' observation.
\item
  \textbf{Stare-mode, synchronised}. Standard `point-and-shoot'
  observations where the mid-point of the UV-VIS and NIR exposures are
  matched.
\item
  \textbf{Nodding-mode}. The telescope `nods' between two-positions
  along the slit throughout exposure, allowing for on-the-fly sky
  background removal.
\item
  \textbf{Fixed sky-offset mode}. This mode is for extended objects
  where not enough uncontaminated sky-background is seen within the 12$''$
  slit to allow for measurement and removal.
\item
  \textbf{Generic sky-offset mode}. User defined pattern of telescope
  offsets.
\item
  \textbf{Mapping-mode}. Used to `map' an object or location.
\end{enumerate}

Together with these spectroscopic modes, an imaging observation mode will also be available via the \ac{AC}. The pipeline will be able to reduced data from this imaging-mode alongside the first 3 spectroscopic modes in a completely automated fashion (see Section \ref{sec:automated_data_reduction}).

Table \ref{tab:products} details each of the final data-products the pipeline shall produce. These products shall meet ESO's Phase III data-products standards and are uploaded and archived on the ESO \ac{SAF} \cite{2011AAS...21830503R} for dissemination to data owners.

\begin{table}[ht]
\caption{Final data-products generated by the SOXS pipeline.} 
\label{tab:products}
\begin{center}       
\begin{tabular}{|l|l|} 
\hline
\rule[-1ex]{0pt}{3.5ex}  Product  & Description         \\
\hline
\rule[-1ex]{0pt}{3.5ex}  1D Source Spectra  & \makecell[l]{1D spectra in FITS binary table format, one for each arm.  \\ Each FITS spectrum file will contain 4 extensions:  \\ 1. Wavelength- and flux-calibrated spectra with absolute flux correction \\ via scaling to acquisition image source photometry, \\ 2. an additional spectrum with correction for telluric absorption via \\ MOLECFIT, \\ 3. the variance array and \\ 4. the sky-background spectra.}      \\
\hline
\rule[-1ex]{0pt}{3.5ex} 1D Merged Source Spectrum & \makecell[l]{1D UV-VIS \& NIR merged spectrum in FITS binary table \\ format with PDF visualisation. This spectrum \\ will be rebinned to a common pixel scale for each arm. \\ This spectrum file will also have the same 4 \\ extensions described above.}  \\
\hline
\rule[-1ex]{0pt}{3.5ex} 2D Source Spectra  & \makecell[l]{A 2D FITS image for each spectral arm containing wavelength \\ and flux calibrated spectra (no other corrections applied) allowing \\ users to perform source extraction with their tool of choice. Note that \\ rectification of the curved orders in the NIR introduces a source \\ of correlated noise not present in extractions performed on the \\ unstraightened orders as done by the pipeline.}   \\
\hline
\rule[-1ex]{0pt}{3.5ex}  Acquisition Camera Images & \makecell[l]{\emph{ugrizy} astrometrically and photometrically (\emph{griz} only) calibrated to \\ Refcat2 \cite{2018ApJ...867..105T} } \\
\hline
\end{tabular}
\end{center}
\end{table} 

\section{Automated Data-Reduction and Data flow}
\label{sec:automated_data_reduction}

The pipeline is being designed to reduce all \ac{SOXS} data to \ac{ESO} Phase III standards `out-of-the-box' with the aim of reducing most of the \ac{SOXS} data and delivering the final-products to the \ac{ESO} \ac{SAF} without the need for human-interaction. The pipeline will run automatically on all point-source targets above an AB magnitude of $r = 19$ (with the stretch goal of $r = 20$). Below this magnitude, the pipeline will \emph{attempt} to run automatically but it may require some user interaction to optimise object extraction. The pipeline may also struggle to provide an optimal absolute flux calibration and/or source extraction for sources with in crowded fields or with complex backgrounds.

The current design for the flow of data from the telescope to the archive is as follows:

\begin{enumerate}
    \item Data is acquired by \ac{NTT} \& \ac{SOXS} on the summit at La Silla, Chile.
\item Raw data is transferred within 10 minutes to the \ac{ESO} \ac{SAF} (Garching, Germany).
\item Raw data is downloaded from the \ac{SAF} to the \ac{SOXS} \ac{DRAWER} cluster in Belfast, UK.
\item Data is reduced on the cluster by the \ac{SOXS} pipeline and streamed back to the \ac{SAF} where it can be accessed by data rights owners.
\end{enumerate}

Our goal is to populate the \ac{ESO} \ac{SAF} with the fully reduced data products within 30 min of the raw exposures appearing in the ESO SAF (15 min stretch goal). This is possible thanks to the fixed format of \ac{SOXS} (apart from the exchangeable slit) allowing calibration frames to be prepared ahead-of-time before science data reductions.

Access to the `open stream' method of shipping reduced data immediately to the \ac{ESO} \ac{SAF} will initially require the \ac{ESO} Quality Control Group to review and verify a moderately sized collection of SOXS-pipeline reduced data. Once the quality and content of the data produced by the pipeline has met with \ac{ESO} Phase III standards, we will then be allowed to upload data-products to the archive without further need of passing through a gatekeeper.

\section{Pipeline Usage}
\label{sec:usage} 

As stated in Section \ref{sec:automated_data_reduction}, the SOXS pipeline will be designed to reduce most SOXS data in a non-interactive, automated fashion. However, users will be able to download and install the pipeline on their own machines and drive the pipeline to reduce the data to their own preferences. We foresee that the pipeline functionality will be accessible via at least two user-interfaces; a \ac{CLI} and via ESO's Reflex GUI. 

The \ac{CLI} being developed has a syntax similar to ESO's \ac{CPL} interface \emph{Esorex} \cite{xshpipe}. Each recipe can be called with an individual command and takes as input a set-of-files (SOF) list containing the required file input (raw data, static calibration products, etc.) and the path to a settings file containing the parameter values for the recipes such as sigma-clipping thresholds, polynomial orders for fitting, etc.

\begin{lstlisting}
Usage:
    soxspipe init
    soxspipe mbias <inputFrames> [-s <pathToSettingsFile>]
    soxspipe mdark <inputFrames> [-s <pathToSettingsFile>]
    soxspipe mflat <inputFrames> [-s <pathToSettingsFile>]
    soxspipe disp_sol <inputFrames> [-s <pathToSettingsFile>]
    soxspipe order_centres <inputFrames> [-s <pathToSettingsFile>]


Options:
    init                                   setup the soxspipe settings file for the first time
    mbias                                  the master bias recipe
    mdark                                  the master dark recipe
    mflat                                  the master flat recipe
    disp_sol                               the disp solution recipe
    order_centres                          the order centres recipe

    inputFrames                            path to a directory of frames or a set-of-files file

    -h, --help                             show this help message
\end{lstlisting}

\begin{figure}[htb!]\centering
  \includegraphics[width=10cm]{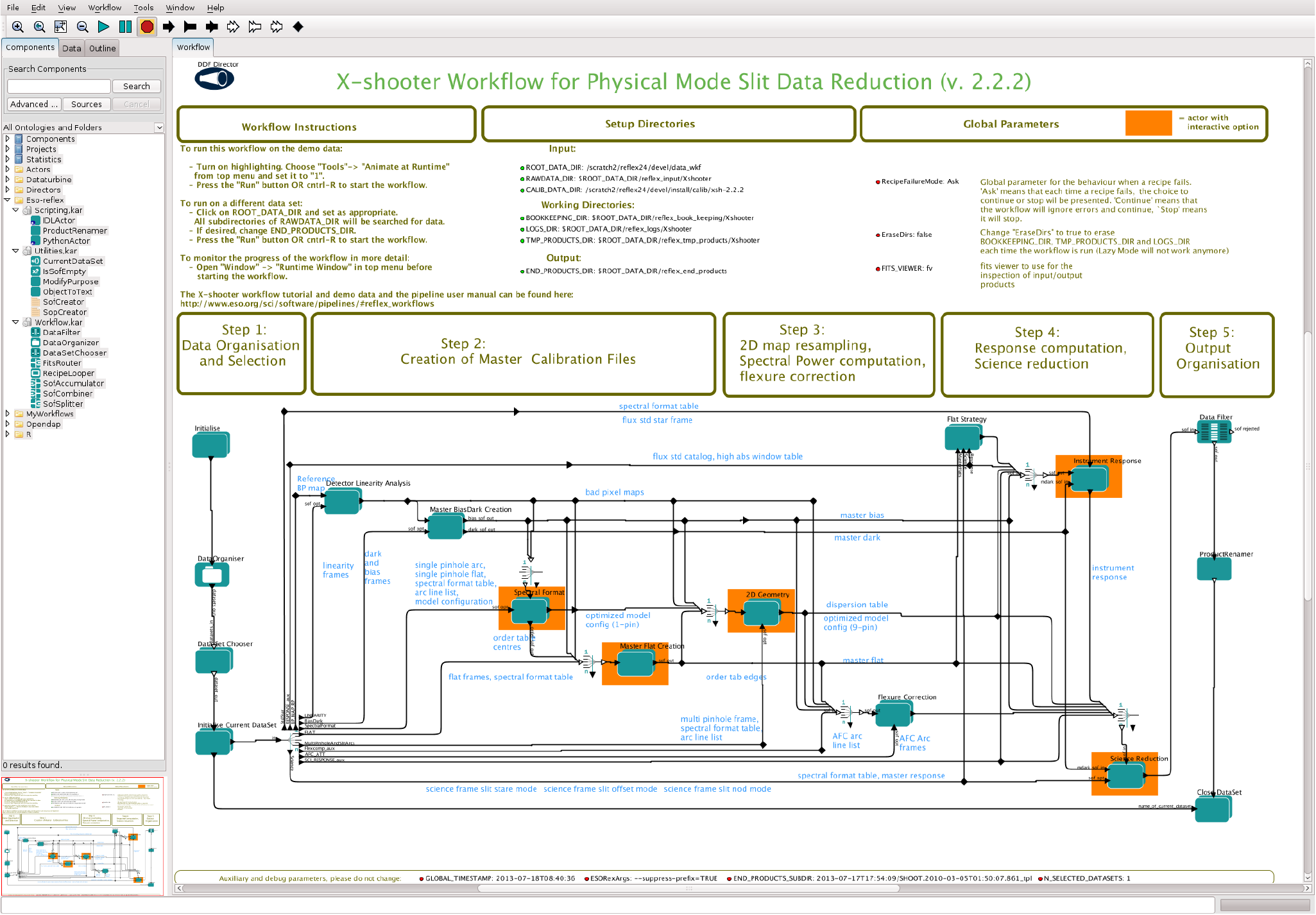}
  \caption{The ESO Reflex environment for X-shooter \cite{reflex}.}
  \label{fig:reflex}
\end{figure}

The most recent version of ESO's \emph{ESOReflex} \cite{reflex} GUI now includes a Python based plugin API that shall allow us to integrate the SOXS pipeline into this data-reduction control and visualisation tool (see Figure \ref{fig:reflex}). Those experienced with the X-Shooter data reductions will be familiar with this tool and should easily transition to reducing SOXS data with the same tool. \emph{ESOReflex} integration will be included in the final version of the pipeline, released after the SOXS's Preliminary Acceptance Chile (PAC).

\section{Development Environment}
\label{sec:development_environment}

At this current point in time the astronomical community have
overwhelmingly adopted Python as their scripting language of choice and
there a plethora of well maintained, mature python packages to help with
basic data-reduction routines not to mention visualisation,
user-interaction and data manipulation. To this end we have opted to
develop the SOXS pipeline in Python 3.

The pipeline is being built with in agile development philosophy that
includes adaptive planning and evolutionary development. As with any
project, one of the greatest risks is knowledge loss due to a team
member leaving the before project completion. To mitigate this risk we
employ pair-programming techniques throughout to share knowledge, both
explicit and tacit, between two developers. In times of travel bans and
remote working a JupterHub server with Python based notebooks and shared
screens within video conferencing tools have been invaluable to execute
these techniques.

We aim for the pipeline to be easy to install and extensively and
clearly documented. Installing the pipeline is as simple as running the
command:

\begin{lstlisting}
pip install soxspipe
\end{lstlisting}

Pipeline documentation is being written in parallel with the code in
docstrings and stand-alone markup files. A push of new code to github
triggers a new build of the documentation on readthedocs\footnote{readthedocs
  https://soxspipe.readthedocs.io/} using the back-end sphinx
documentation engine.

\subsection{Test Driven Development}

To verify the pipeline is not only able to reduce a typical data-set but
also data that is far from ideal we are using test driven development
throughout the build using `extreme' mock data. This data helps push the
pipeline to the limits of it capabilities and allows us to develop for
possible edge-case scenarios that the pipeline will experience in
real-time production.

\begin{figure}[htb!]\centering
  \includegraphics[width=10cm]{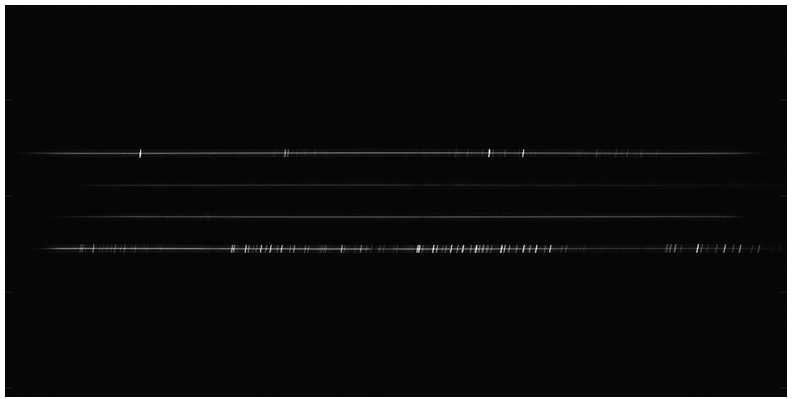}
  \caption{Simulated UV-VIS spectral image of a GOV star ($V=18$)  object and different tilts on the slit for each pseudo-order. Seeing 0.87$''$, slit 1$''$, exposure 1800s.}
  \label{fig:uv_vis_model_with_a_bright_source}
\end{figure}

\begin{figure}[htb!]\centering
  \includegraphics[width=10cm]{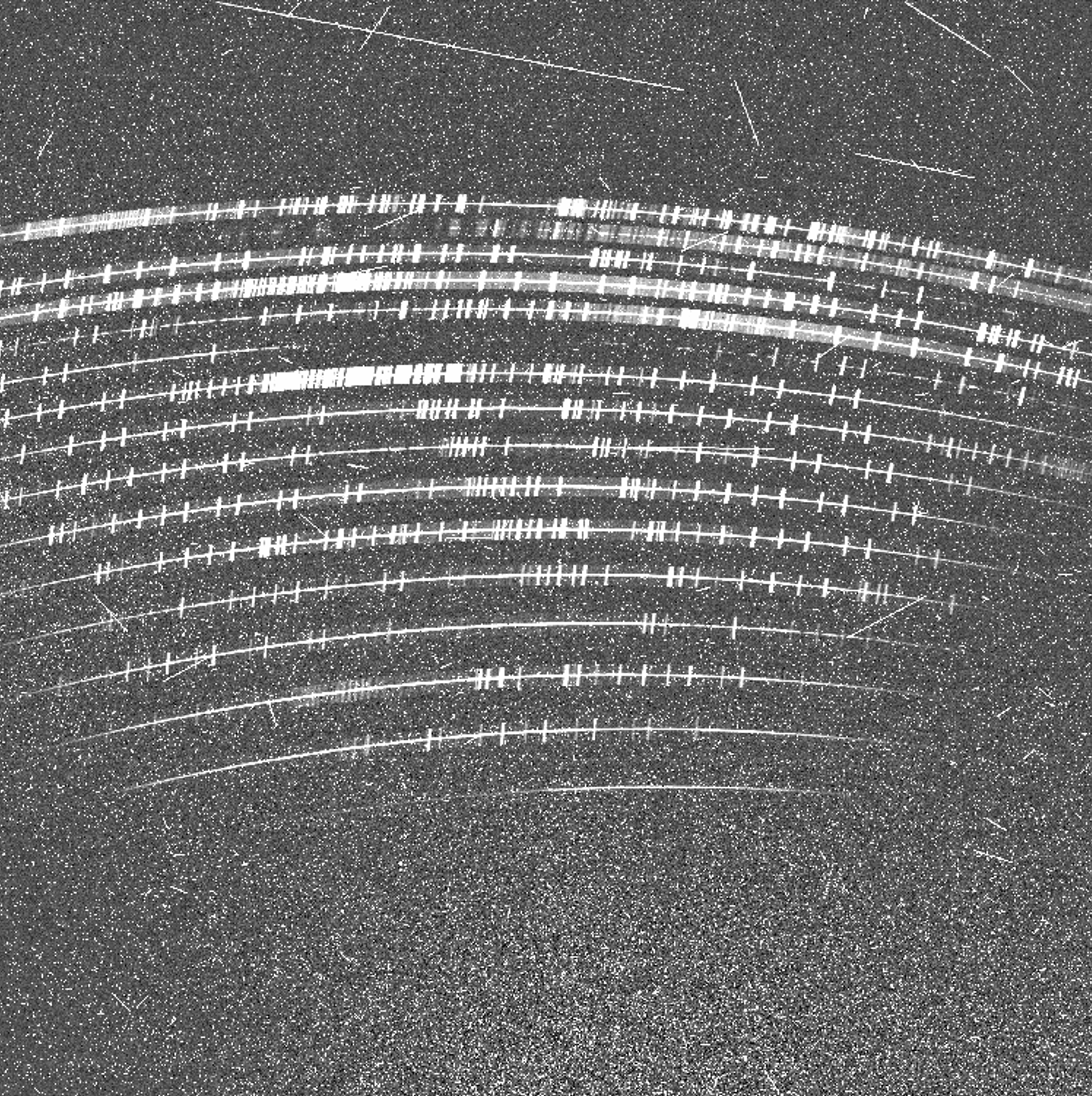}
  \caption{Full NIR arm displaying a synthetic image of black body at 5800 K at $V=16$ mag point-source. Seeing 1$''$, slit 1$''$, 300 sec exposure.}
  \label{fig:nir_model_with_point_source}
\end{figure}

This mock data is generated via the SOXS End-to-End (E2E) simulator
\cite{soxsgenoni} which is able to synthesise 2D images that take into
account the main optical behaviour of the system (grating dispersion,
sampling, PSF, noises and position of various resolution elements coming
from full ray-tracing). An example of the mock data is reported in
Figures \ref{fig:uv_vis_model_with_a_bright_source} and \ref{fig:nir_model_with_point_source}.

In particular, the plan is to generate mock data by varying different
observing conditions in terms of source magnitude (faint or bright
continuum), sky-brightness and calibration frames so as to provide a
full suite of stress tests for point-like sources down to the limit
magnitude considered for SOXS.

The E2E model will be also able to provide synthetic simulated frames
produced by the acquisition camera with the same aim to test the data
flow of the reduction chain for the photometry performed with the
\ac{AC}.

Code is version controlled with git, hosted on github\footnote{github
  https://github.com/thespacedoctor/soxspipe} and linked to a Jenkins
Continuous Integration/Continuous Deployment server via webhooks. A push
of new code to any branch in the github pipeline repository triggers a
new `build' of the code where all unit-tests are ran. If all tests pass
then the branch is merged into its parent branch (which in turn triggers
a new testing of the parent branch). If the master branch being tested
and all tests pass then a new dot release version of the code is shipped
to PyPI\footnote{pypi https://pypi.org/project/soxspipe/} for
deployment.

\section{Pipeline Architecture}
\label{sec:architecture} 

The SOXS pipeline borrows the informative concept of `recipes' employed by ESO's \ac{CPL} to define the modular components of the pipeline. These recipes can be strung together to create an end-to-end workflow that takes as input the raw and calibration frames from the instrument and telescope and processes them all the way through to fully reduced, calibrated, ESO Phase III compliant science products.

We also employ the term `utilities' to define reusable functions designed to be called from multiple recipes. Recipes are named with the prefix `soxs' (e.g. \texttt{soxs\_mbais}) followed by a succinct description of the recipe.

\begin{figure}[h]\centering
  \includegraphics[width=15cm]{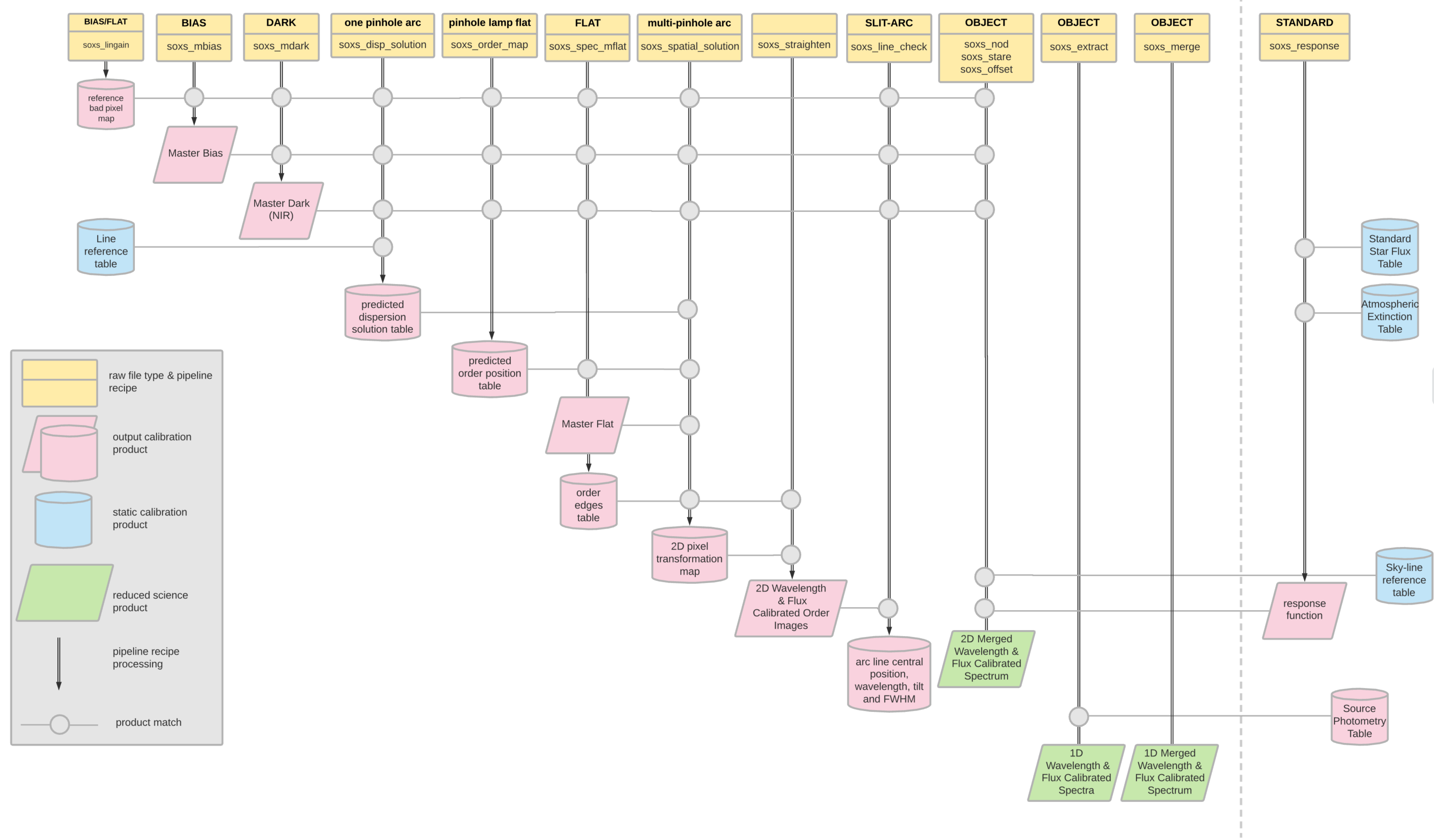}
  \caption{The SOXS Spectroscopic Data Reduction Cascade}
  \label{fig:soxs_spectroscopic_data_reduction_cascade}
\end{figure}

\begin{figure}[h]\centering
  \includegraphics[width=10cm]{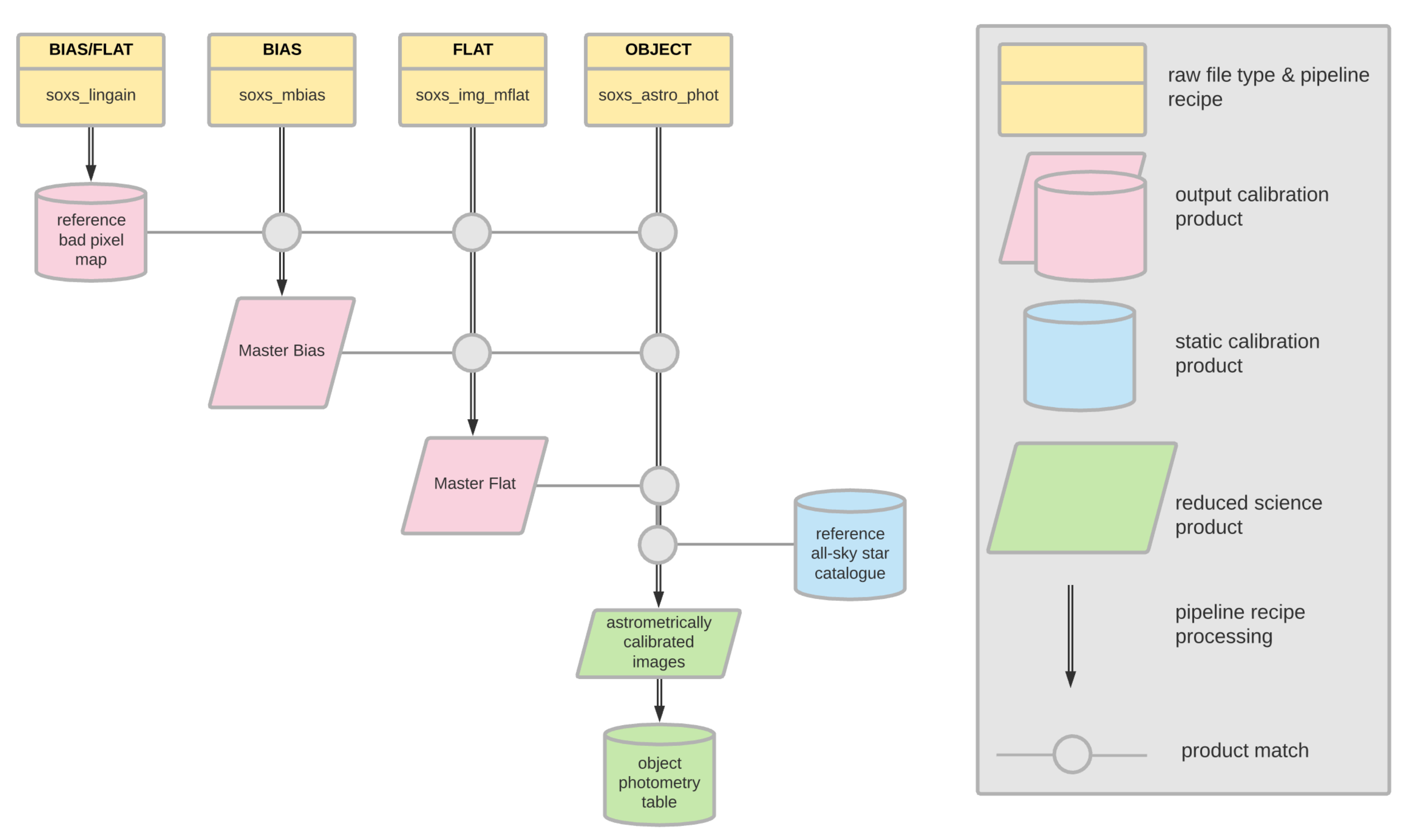}
  \caption{SOXS Imaging Data Reduction Cascade}
  \label{fig:soxs_imaging_data_reduction_cascade}
\end{figure}

\begin{table}[ht]
\caption{Index of SOXS Pipeline Recipes.} 
\label{tab:recipe-index}
\begin{center} 
\begin{tabular}{|l|l|l|}
\hline
\rule[-1ex]{0pt}{3.5ex}   Recipe & Reduction Stage  \\ \hline
\rule[-1ex]{0pt}{3.5ex}  \texttt{soxs\_data\_organiser} & \cellcolor{blue} pre-processing   \\ \hline
\rule[-1ex]{0pt}{3.5ex}  \texttt{soxs\_lingain} & \cellcolor{violet} calibration  \\ \hline
\rule[-1ex]{0pt}{3.5ex}  \texttt{soxs\_img\_mflat} & \cellcolor{violet} calibration  \\ \hline
\rule[-1ex]{0pt}{3.5ex}  \texttt{soxs\_mbias} & \cellcolor{violet} calibration  \\ \hline
\rule[-1ex]{0pt}{3.5ex}  \texttt{soxs\_mdark} & \cellcolor{violet} calibration \\ \hline
\rule[-1ex]{0pt}{3.5ex}  \texttt{soxs\_disp\_solution} & \cellcolor{green} rectification  \\ \hline
\rule[-1ex]{0pt}{3.5ex}  \texttt{soxs\_order\_centres} & \cellcolor{green} rectification  \\ \hline
\rule[-1ex]{0pt}{3.5ex}  \texttt{soxs\_spatial\_solution} & \cellcolor{green} rectification  \\ \hline
\rule[-1ex]{0pt}{3.5ex}  \texttt{soxs\_spec\_mflat} & \cellcolor{green} rectification  \\ \hline
\rule[-1ex]{0pt}{3.5ex}  \texttt{soxs\_straighten} & \cellcolor{green} rectification  \\ \hline
\rule[-1ex]{0pt}{3.5ex}  \texttt{soxs\_line\_check} & \cellcolor{green} rectification  \\ \hline
\rule[-1ex]{0pt}{3.5ex}  \texttt{soxs\_nod} & \cellcolor{orange} sky-subtraction  \\ \hline
\rule[-1ex]{0pt}{3.5ex}  \texttt{soxs\_stare} & \cellcolor{orange} sky-subtraction  \\ \hline
\rule[-1ex]{0pt}{3.5ex}  \texttt{soxs\_offset} & \cellcolor{orange} sky-subtraction  \\ \hline
\rule[-1ex]{0pt}{3.5ex}  \texttt{soxs\_extract} & \cellcolor{red} extraction+   \\ \hline
\rule[-1ex]{0pt}{3.5ex}  \texttt{soxs\_response} & \cellcolor{red} extraction+   \\ \hline
\rule[-1ex]{0pt}{3.5ex}  \texttt{soxs\_merge} & \cellcolor{red} extraction+   \\ \hline
\rule[-1ex]{0pt}{3.5ex}  \texttt{soxs\_astro\_phot} & \cellcolor{red} extraction+   \\ \hline
\end{tabular}
\end{center} 
\end{table}
The reduction cascade of SOXS pipeline for spectroscopic data is reported in Figure \ref{fig:soxs_spectroscopic_data_reduction_cascade} while the cascade associated to acquisition camera imaging reduction is depicted in Figure \ref{fig:soxs_imaging_data_reduction_cascade}.

In particular, the association maps in Figures \ref{fig:soxs_spectroscopic_data_reduction_cascade} and \ref{fig:soxs_imaging_data_reduction_cascade} show how spectral and imaging data cascade flow through the SOXS pipeline. The input data, calibration products required and the output frames are shown for each pipeline recipe implemented in the pipeline. Each of the vertical lines in the map depict a raw data frame, the specific recipe to applied to that frame and the data product(s) output by that recipe. Horizontal lines show how those output data products are used by subsequent pipeline recipes. Time loosely proceeds from left-to-right (recipe order) and from top-to-bottom (recipe processing steps) on the map. To the right of the grey dashed line in \ref{fig:soxs_spectroscopic_data_reduction_cascade} there are input calibration products generated from a separate pipeline processing cascade.

\begin{figure}[h]\centering
  \includegraphics[width=8cm]{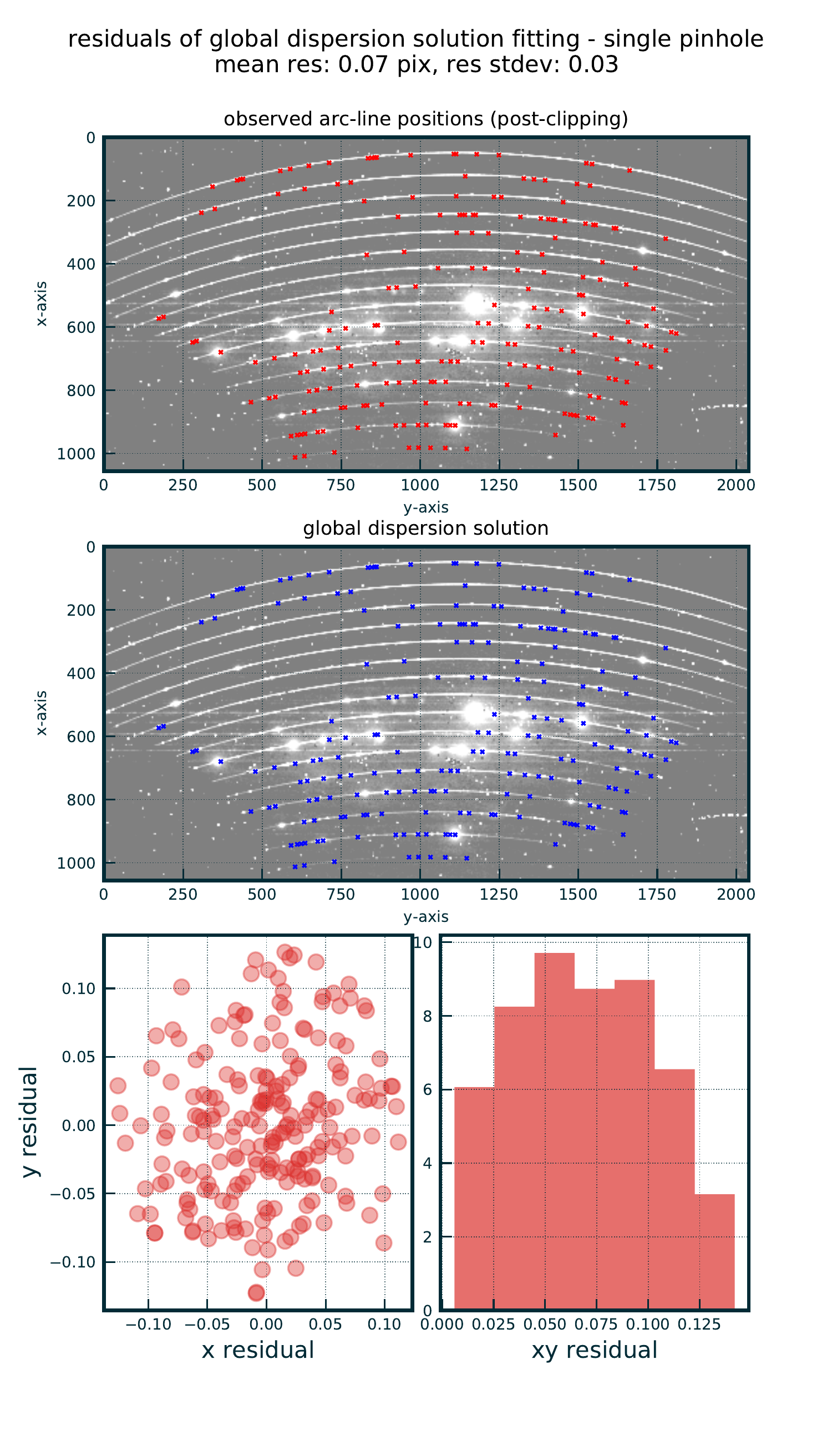}
  \caption{\texttt{soxs\_disp\_solution} recipe output. Top panel - Identified lines against a raw frame single pinhole illuminated by calibration lamp. Central panel - Survived spectral lines after sigma clipping and iteration on the global fit. Bottom panel: residuals (in px)}.
  \label{fig:soxs_dispsolution}
\end{figure}

\begin{figure}[h]\centering
  \includegraphics[width=8cm]{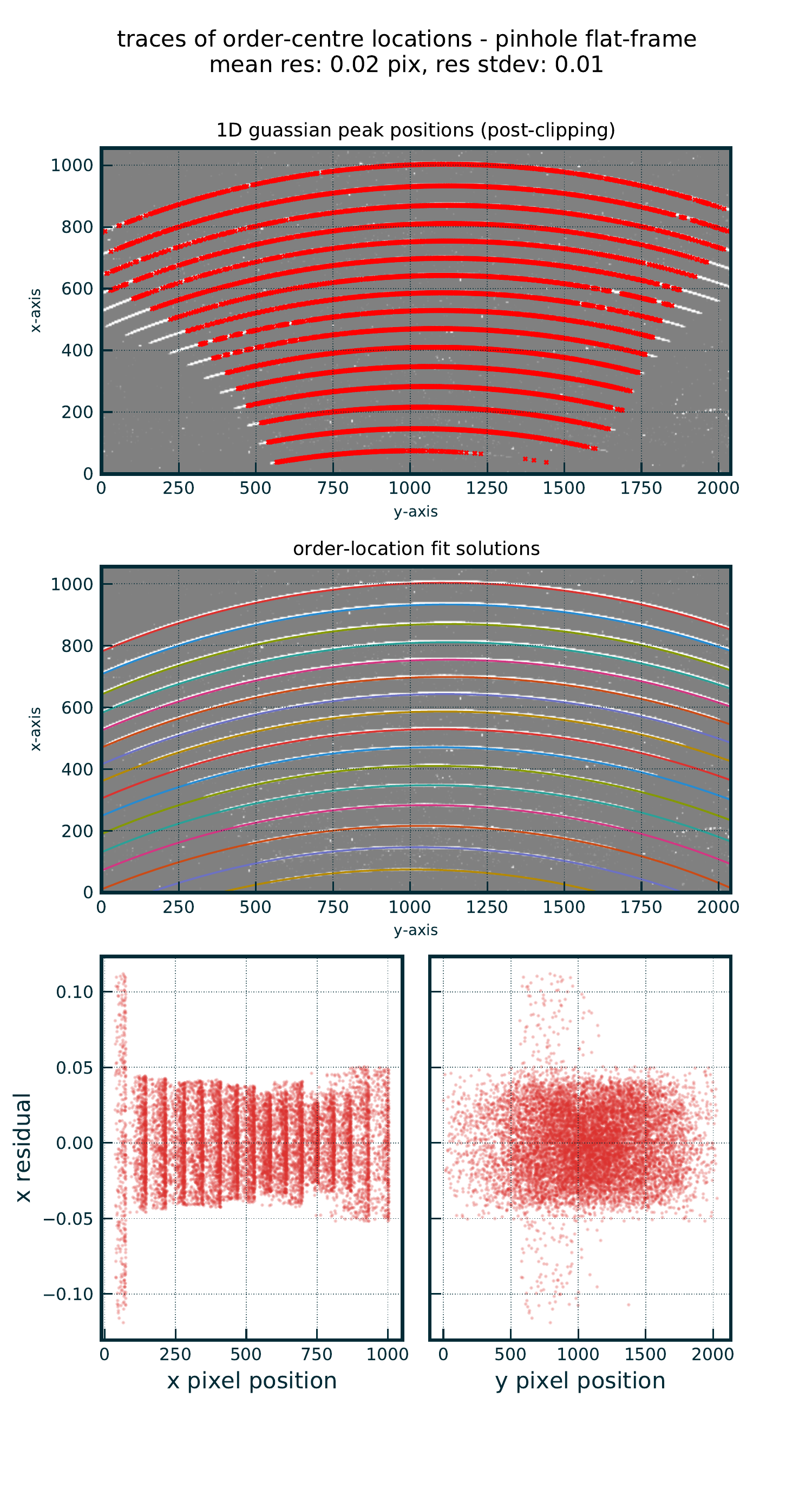}
  \caption{\texttt{soxs\_order\_centres} recipe output. Top panel - Fitted position of orders on the raw single-pinhole image. Central panel - Global fit obtained by interpolating red points of top panel. Bottom panel: residuals (in px)}.
  \label{fig:soxs_oc}
\end{figure}

\begin{figure}[h]\centering
  \includegraphics[width=8cm]{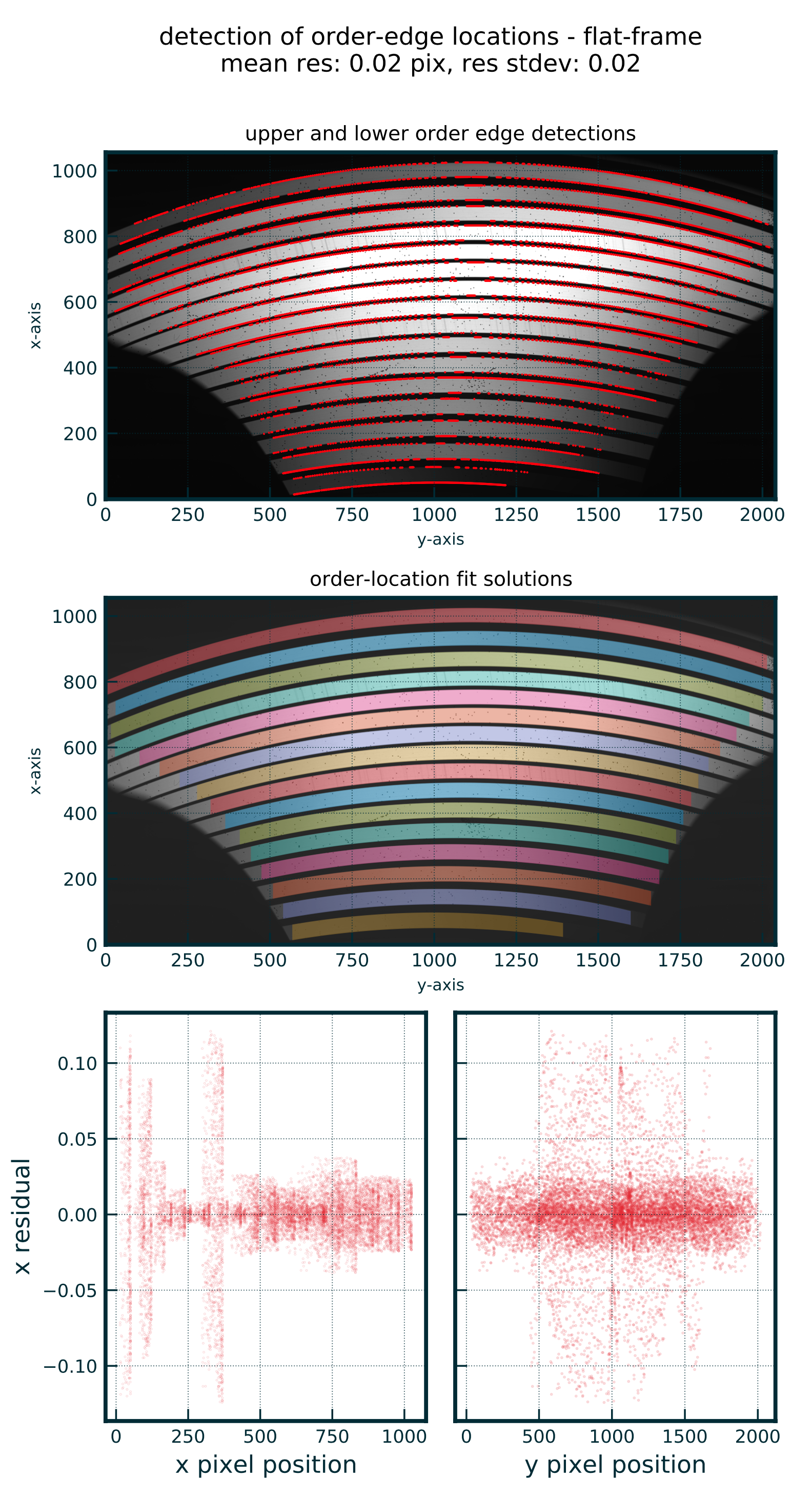}
  \caption{\texttt{soxs\_mflat} recipe output. Top, Central and bottom panel are similar to Figure \ref{fig:soxs_oc}}.
  \label{fig:soxs_flat}
\end{figure}

The recipes required to reduce spectroscopic data from raw frames to 1D flux calibrated spectra are:
\begin{itemize}
    \item {\texttt{soxs\_lingain}: Identify pixels that respond to varying levels of flux in a different way compared to the typical pixel.}
    \item {\texttt{soxs\_mbias}: Used to create Master Bias frame and the first guess of bad pixels on the detectors.}
    \item {\texttt{soxs\_mdark}: Remove the mean dark current from images and identify all hot or cold pixels.}
    \item {\texttt{soxs\_disp\_solution}: Generate first guess of the instruments spectral format and dispersion solutions and central trace positions for each order. Results from this recipe are shown in Figure \ref{fig:soxs_dispsolution}}.
    \item {\texttt{soxs\_order\_centres}: This recipe, starting from the traces first approximated in the \texttt{soxs\_disp\_solution} accurately measures the central trace of each order. An example of the output of the recipe and its ac hived performance is reported in Figure \ref{fig:soxs_oc}}.
    
    \item {\texttt{soxs\_spec\_mflat}: The recipe generates master flat-field frames, flag hot and dead pixels. The spatial extension of the orders is also measured at this stage. Figure \ref{fig:soxs_flat} depicts results obtained against X-shooter NIR arm data. }
    
    \item{\texttt{soxs\_spatial\_solution}: This recipe estimates the spatial- \textit{and} dispersion-solutions to generate a 2D map used later to translate curved order images into a linear ($\lambda$, s) space. The recipe is similar in logic to \texttt{soxs\_disp\_solution} but it also samples along the slit in the cross-dispersion direction (i.e. Y-axis) by using the multi-pinhole slit frames illuminated by the arc lamp.} 
    
    \item{\texttt{soxs\_straighten}: This recipe uses the 2D map (modeled via polynomials) computed in \\ \texttt{soxs\_spatial\_solution} to straighten (rectify) the curve order images and merge them together. Users will be able (if they wish) to perform source extraction on these rectified frames via the pipeline or using their own extraction routines.} 
    
    \item{\texttt{soxs\_nod, soxs\_stare, soxs\_offset}: These recipes are in charge of extracting the spectra in counts (order by order) and, in the case of stare mode, produce a model of the sky spectrum to be subtracted before extraction of the science pixels.}
    
    \item{\texttt{soxs\_response}: Starting from observation of standard stars, this recipe computes the efficiency curve and response curve of SOXS to flux calibrate spectra for each order.}
    
    \item{\texttt{soxs\_merge}: The orders extracted and flux calibrated in the previous step are merged together in order to obtain a unique, rectified and calibrated 1D spectra of the source (with sky subtraction).}
    
\end{itemize}

\section{Conclusions}
\label{sec:conclusions} 

SOXS is expected to be delivered to Chile (La Silla) in mid 2022 and offered to the community for the GTO and regular proposals from ESO by the end 2022. To meet this goal, we designed an agile framework of development based on Python 3 to develop a robust pipeline that can be used both automatically and interactively by supporting standard Esoreflex workflow.
\\In the paper, we presented the current status of the  pipeline by especially focusing on the design architecture and development environment that foresee the use of current state-of-the-art technologies in the context of software development frameworks. 
At the moment of writing of this manuscript, the team is almost on schedule for delivering the pipeline by end 2021 where first tests of SOXS will take place in Europe. The adoption of agile framework, presented in Section 6 and the use of Python 3 greatly helped on meeting the time constraint while assuring high quality of the written software in terms of  maintainability, readability and robustness.

\acknowledgments 
 
\bibliography{report} 
\bibliographystyle{spiebib} 

\end{document}